\documentclass[aps,twocolumn,showpacs,lengthcheck,preprintnumbers]{revtex4-1}
\usepackage{amsfonts}
\usepackage{amsmath}
\usepackage{amssymb}
\usepackage{graphicx}
\usepackage{fontenc}
\usepackage{color}

\begin{document}
\title{Stripe and junction-vortex phases in linearly coupled Bose-Einstein condensates}
	
\author{Haibo Qiu}
\affiliation{
	School of Science, Xi’an University of Posts and Telecommunications, Xi’an 710121, China
}
\author{Dengling Zhang}
\affiliation{
	School of Science, Xi’an University of Posts and Telecommunications, Xi’an 710121, China
}
\author{Antonio Mu\~{n}oz Mateo}
\affiliation{Departamento de F\'isica, Universidad de La Laguna, La Laguna, Tenerife 38200, Spain}


\begin{abstract} 
Soon after its theoretical prediction, striped-density states in the presence of synthetic spin-orbit coupling were realized in Bose-Einstein condensates of ultracold neutral atoms [J.-R. Li et al., Nature \textbf{543}, 91 (2017)]. The achievement opens avenues to explore the interplay of superfluidity and crystalline order in the search for supersolid features and materials. The system considered is essentially made of two linearly coupled Bose-Einstein condensates, that is a pseudo-spin-$1/2$ system, subject to a spin-dependent gauge field $\sigma_z \hbar k_\ell$. Under these conditions the stripe phase is achieved when the linear coupling $\hbar\Omega/2$ is small against the gauge energy $m\Omega/\hbar k_\ell^2<1$ . The resulting density stripes have been interpreted as a standing-wave, interference pattern with approximate wavenumber $2k_\ell$.  
Here, we show that the emergence of the stripe phase is induced by an array of Josephson vortices living in the junction defined by the linear coupling. As happens in superconducting junctions subject to external magnetic fields, a vortex array is the natural response of the superfluid system to the presence of a gauge field. Also similar to superconductors, the Josephson currents and their associated vortices can be present as a metastable state in the absence of gauge field. We provide closed-form solutions to the 1D mean field equations that account for such vortex arrays. The underlying Josephson currents coincide with the analytical solutions to the sine-Gordon equation for the relative phase of superconducting junctions [C. Owen and D. Scalapino, Phys. Rev. \textbf{164}, 538 (1967)].
\end{abstract}

\maketitle

\section{Introduction}
The interplay of crystalline order and Bose-Einstein condensation encompasses relevant physical phenomena from the realm of superconductors, where it rules the motion of Cooper pairs of fermions through crystals \cite{Plakida2012}, to the field of ultracold atomic gases, where laser-induced lattices modulate the superfluid flow of condensed bosons  \cite{Morsch2006}, and also up to the emergence of the supersolid phase \cite{Andreev1969,Chester1970,Legget1970}.
Supersolidity, by means of which crystalline structures are endowed with superfluid properties, was envisaged to take place in solid helium at very low temperature and very high pressure \cite{Chan2013}. 
However, only recently, supersolid properties have been realized in ultracold-gas systems by means of spin-orbit coupling \cite{Li2017}, by using Bose-Einstein condensate (BEC) to optical-cavity coupling \cite{Leonard2017,Leonard2017b}, and also as a result of competing contact and dipolar interactions \cite{Bottcher2019,Chomaz2019}. 
These diverse methods have been used to simultaneously achieve a phase coherence, capable of providing the superfluid property, and a modulated particle density, reflecting the translational symmetry breaking of crystals. 

 The present paper focuses on the first of these arrangements, using a gauge field to obtain a modulated density, but it detours from the pure supersolid, density features of the system  (see, e.g., Ref. \cite{Li2013} in this regard), to aim at the study of its phase properties, viewed as the ultimate cause of the observed density modulation. 

The accomplishment of spin-dependent synthetic gauges in atomic gases that are electrically neutral has represented a breakthrough for the simulation of quantum systems \cite{Lin2011}. 
The spin-orbit-coupled configuration arranged in Ref. \cite{Li2017} to observe the predicted stripe phase \cite{Ho2011,Li2012a,Li2015} is a relevant instance, where the spin-dependent gauge, in combination with the linear coupling of a  pseudo-spin-$1/2$ BEC, simulates the dynamics of a superconducting junction in the presence of an external magnetic field. In superconductors, Josephson currents are expected to flow due to the relative phases induced by the vector potential.
 In the regime where the penetration of the magnetic field in the junction (the Josephson penetration depth) is smaller than the junction extension, the junction response mimics a weak-superconductor bulk and is only threaded by singular flux lines that define Josephson vortices \cite{Barone1982}. 
 Such a response was analytically calculated in one-dimensional (1D) junctions, with and without external magnetic field, by Owen and Scalapino in 1967 as solutions to the Sine-Gordon equation \cite{Owen1967} (and also revealed that same year by P. W. Anderson \cite{Anderson1967}). 
If the Josephson junction of the analog pseudo-spin-$1/2$ BEC is equivalently long, which is ruled by the linear coupling, the ultracold-gas response to a gauge field should follow the same dynamics as the simulated superconducting system. In the rest of the paper we show that this is indeed the case, and the bosonic junction become threaded by an array of vortices whose positions mark the minima of the density stripes in the bulk condensate.

The junction vortices can equally emerge in the absence of gauge field whenever a chiral current density flows around the junction. In this regard, recent experiments in bosonic ultracold gases loaded in double-well potentials have detected sine-Gordon solitons in the relative phase of the linearly coupled atomic clouds localized at each well \cite{Schweigler2017}. Such solitons, as it will become apparent later, are the fingerprint of the underlying, localized chiral currents that give rise to Josephson vortices.

\section{Pseudo-spin-$1/2$ BEC}
The BEC dynamics will be modeled by the Gross-Pitaevskii (GP) equation for the pseudo-spin-$1/2$ order parameter $\Psi=[\Psi_\uparrow \;\Psi_\downarrow]^T$
\begin{align}
i\hbar\frac{\partial\Psi}{\partial t}  =
\left(\begin{array}{cc} 
\frac{\hat p^2}{2m}+g \left\vert \Psi_{\uparrow}\right\vert ^{2}& -\frac{\hbar\Omega}{2} \\
-\frac{\hbar\Omega}{2} & \frac{\hat p^2}{2m}+g \left\vert \Psi_{\downarrow}\right\vert ^{2}
\end{array} \right) \Psi,
\label{eq:gp}
\end{align}
where $\hat p=-i\hbar\partial_x$ is the momentum operator, and the strength of the contact, repulsive interparticle interaction $g=4\pi\hbar^2a_s/m$ is measured by the positive $s$-wave scattering length $a_s>0$. In the presence of spin-orbit-coupling, the momentum operator transforms as $\hat p \rightarrow \hat p + \sigma_z \hbar k_\ell $, by adding the momentum contribution of the gauge field $\hbar k_\ell$ through the Pauli matrix $\sigma_z$. For the sake of simplification of the analytical treatment, we will assume periodic boundary conditions in the spatial coordinate $\Psi_\sigma(x,t)=\Psi_\sigma(x+L,t)$, with $\sigma=\uparrow,\, \downarrow$ , that is, a 1D ring geometry of length $L$. 
The normalization is fixed by the number of particles $N=\int_L dx\,(|\Psi_{\uparrow}|^2+|\Psi_{\downarrow}|^2)$, which is a conserved quantity derived from the continuity equations $\partial_t |\Psi_{\uparrow}|^2+ \partial_x J_{\uparrow} =-(\partial_t |\Psi_{\downarrow}|^2+\partial_x J_{\downarrow})=\mathcal{J}_\Omega$. Here $J_{\sigma}=|\Psi_{\sigma}|^2 \mathrm{v}_\sigma$ are the axial particle current densities in the condensates, with superfluid velocities $\mathrm{v}_{\sigma}=(\hbar/m)\partial_x \arg \Psi_{\sigma}$, whereas $\mathcal{J}_\Omega=\Omega |\Psi_{\uparrow}| |\Psi_{\downarrow}|\,\sin\varphi$ is the Josephson current of tunneling particles
across the junction, which is modulated by the relative phase $\varphi(x)= \arg \Psi_{\uparrow}- \arg \Psi_{\downarrow}$. In the presence of spin-orbit-coupling, the continuity equations hold with shifted superfluid velocities $\mathrm{v}=(\hbar/m)(\partial_x \arg \Psi +\sigma_z k_\ell)$.

\section{ Generalized sine-Gordon equation}
The symmetry of Eq. (\ref{eq:gp}) suggests the search for stationary states with sharing density profiles $|\Psi_{\uparrow}|^2=|\Psi_{\downarrow}|^2=n(x)$ and opposite superfluid velocities $\mathrm{v}_{\downarrow}=-\mathrm{v}_{\uparrow}$. In this case, by subtracting the two continuity equations, one gets a single equation for the relative phase that rules the particle currents:
\begin{align}
\frac{\partial}{\partial x}\left[n(x)\frac{\partial \varphi}{\partial x} \right]=\frac{n(x)}{\lambda_\Omega^2}\sin \varphi,
\label{eq:sineG}
\end{align}
where $\lambda_\Omega=\sqrt{\hbar/2m\Omega}$ is the characteristic length determined by the linear coupling. The square parenthesis is intrinsically the chiral density current $J_{\uparrow}-J_{\downarrow}=(\hbar/m)\,n(x)\partial_x \varphi$, which 
  provides the superfluid relative velocity $\mathrm{v}_{\uparrow}-\mathrm{v}_{\downarrow}=(J_{\uparrow}-J_{\downarrow})/n(x)$. The presence of spin-orbit coupling transforms just the chiral density current in Eq. (\ref{eq:sineG}) such that $J_{\uparrow}-J_{\downarrow} = (\hbar/m)\,n(x)(\partial_x \varphi+2k_\ell)$. 
  By making use of the gauge invariance property, one can transform the phase $\varphi\rightarrow \varphi'=\varphi+\chi$ and simultaneously the gauge field $2 k_\ell\rightarrow 2 k_\ell'=2 k_\ell-\partial_{x}\chi$, so that the particle currents remain unchanged. In particular, by using $\chi=2 k_\ell x$ the gauge field is moved from the chiral current to the Josephson current, which in the new gauge becomes $\mathcal{J}_\Omega=\Omega |\Psi_{\uparrow}| |\Psi_{\downarrow}|\,\sin (\varphi'-2 k_\ell x)$.

Equation (\ref{eq:sineG}) reduces to the stationary sine-Gordon equation when the condensate density is constant $ n(x)=\bar n$, which is the general assumption inside the superconducting junctions. There, the sine-Gordon equation is obtained from the electrodynamic relation between the junction current $\mathcal{J}_J$ and the total (externally applied plus current-induced) magnetic field $\partial_x H_y=\mathcal{J}_J $, where, for definiteness, we have assumed a $y$-oriented field in a planar $xy$ junction with long axis $x$ and unit length across. This Maxwell equation is accompanied by the relation between the variation of the gauge-invariant phase $\varphi_{sc}$ and the magnetic flux, whereby $\partial_x\varphi_{sc}=(2\pi d/\Phi_o) H_y$,
where $d$ is a characteristic transverse length and $\Phi_o$ is the flux quantum \cite{Barone1982}. From this latter expression, and inspection of Eq. (\ref{eq:sineG}), it is clear that, apart from constants, the chiral current in the neutral gas plays the role of the total magnetic field threading the superconducting junction, and therefore it determines the junction dynamics. In particular, from comparison of both mathematical models, by writing $(m/\hbar)(J_{\downarrow}-J_{\uparrow})/\bar n=\alpha\partial_x \varphi_{sc}$ and $(n(x)/\bar n) \sin\varphi=\alpha\sin\varphi_{sc}$, with $\bar n$ a constant density and $\alpha$ a non-dimensional constant, one recovers an equivalent sine-Gordon equation in the new relative phase $\varphi_{sc}$. 
Although it is not apparent that this mapping can be realized, as we show next, this is indeed the case in the absence of spin-orbit coupling. From direct integration, Owen and Scalapino found two types of general solutions to the 1D sine-Gordon equations \cite{Owen1967},  namely $\varphi_{sc,0}(x)=\arcsin[2\, \mathrm{sn}(kx,\mathfrak{m})\, \mathrm{dn}(kx,\mathfrak{m})]$, and $\varphi_{sc,1}(x)=\arcsin[2\, \mathrm{sn}(kx,\mathfrak{m})\, \mathrm{cn}(kx,\mathfrak{m})]$, where  $\mathrm{sn},\,\mathrm{cn},\,\mathrm{dn}$ stands for the Jacobi elliptic functions \cite{NIST:DLMF} of parameter $\mathfrak{m}\in[0,1]$, and "wavenumber" $k=2jK(\mathfrak{m})/L$, with $j=2,\,4,\,6...$ (in a system with periodic boundary conditions), and $K(\mathfrak{m}) \in [\pi/2,\,+\infty)$ is the complete elliptic integral of the first kind. 
By mapping these solutions into the neutral gas model, and despite the fact that the corresponding phases $\varphi_{0}(x)$ and $\varphi_{1}(x)$ also depend on yet unknown condensate densities $n_0(x)$ and $n_1(x)$, one gets general expressions for the axial current densities and the particle tunneling across the junction:
\begin{align}
\varphi_{0}(x)=\arcsin\left[2\,\bar n\,\alpha \frac{\mathrm{sn}(kx,\mathfrak{m})\, \mathrm{dn}(kx,\mathfrak{m})}{n_0(x)}\right],
\label{eq:JV0_phase}
\\
J_{0,\downarrow}=-J_{0,\uparrow}= \frac{\hbar k}{m}\bar n\,\alpha\, \mathrm{cn}(kx,\mathfrak{m}),
\label{eq:JV0_current}
\\
\mathcal{J}_{0,\Omega}= 2\,\Omega \bar n \alpha \, \mathrm{sn}(kx,\mathfrak{m})\, \mathrm{dn}(kx,\mathfrak{m}),
\label{eq:JV0_tunnel}
\end{align}
and
\begin{align}
\varphi_{1}(x)=\arcsin\left[2\,\bar n\,\alpha \frac{\mathrm{sn}(kx,\mathfrak{m})\, \mathrm{cn}(kx,\mathfrak{m})}{n_1(x)}\right],
\label{eq:JV1_phase}
\\
J_{1,\downarrow}=-J_{1,\uparrow}= \frac{\hbar k}{m}\bar n\,\alpha \,\mathrm{dn}(kx,\mathfrak{m}),
\label{eq:JV1_current}
\\
\mathcal{J}_{1,\Omega}= 2\,\Omega\, \bar n\, \alpha \, \mathrm{sn}(kx,\mathfrak{m})\, \mathrm{cn}(kx,\mathfrak{m}),
\label{eq:JV1_tunnel}
\end{align}
where due to the constraints posed by the continuity equations $\hbar k^2/2m=\Omega$ for $\varphi_{0}$, and  $\mathfrak{m}\,\hbar k^2/2m=\Omega$ for $\varphi_{1}$.
\begin{figure*}[tb]
	\flushleft (\textbf{a})  \hspace{\columnwidth}(\textbf{b}) \\
	\vspace{-0.5cm}
	\centering
	\includegraphics[width=0.4\linewidth]{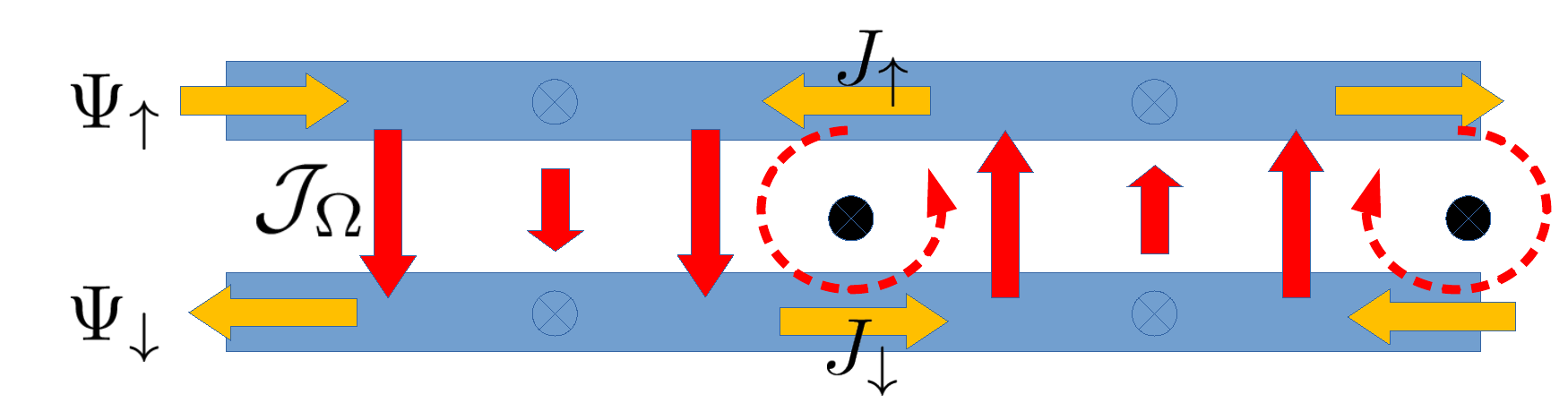}\hspace{1.4cm}
	\includegraphics[width=0.4\linewidth]{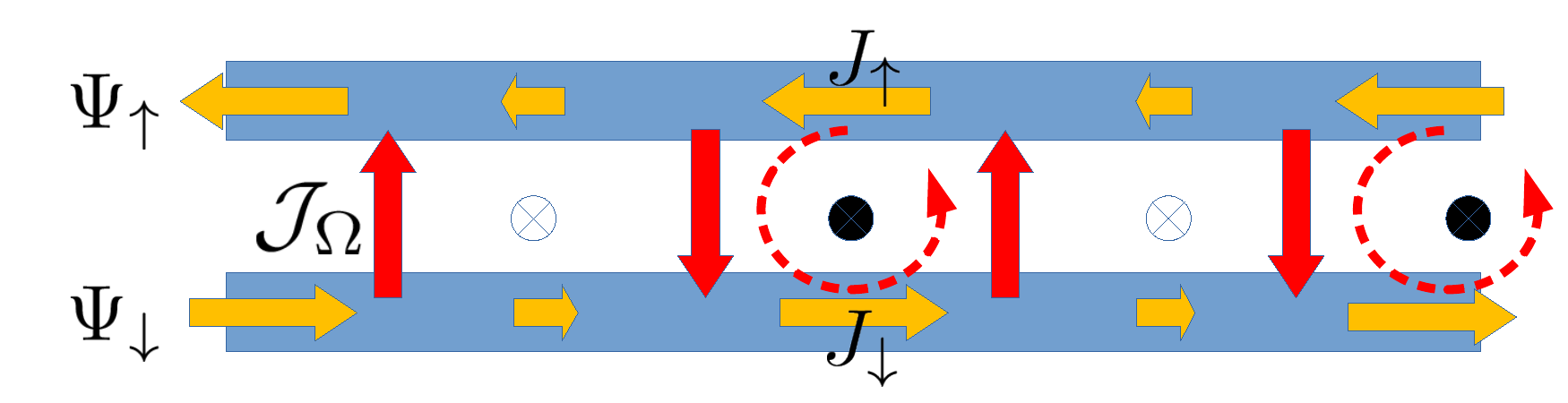}\\
	\includegraphics[width=0.4\linewidth]{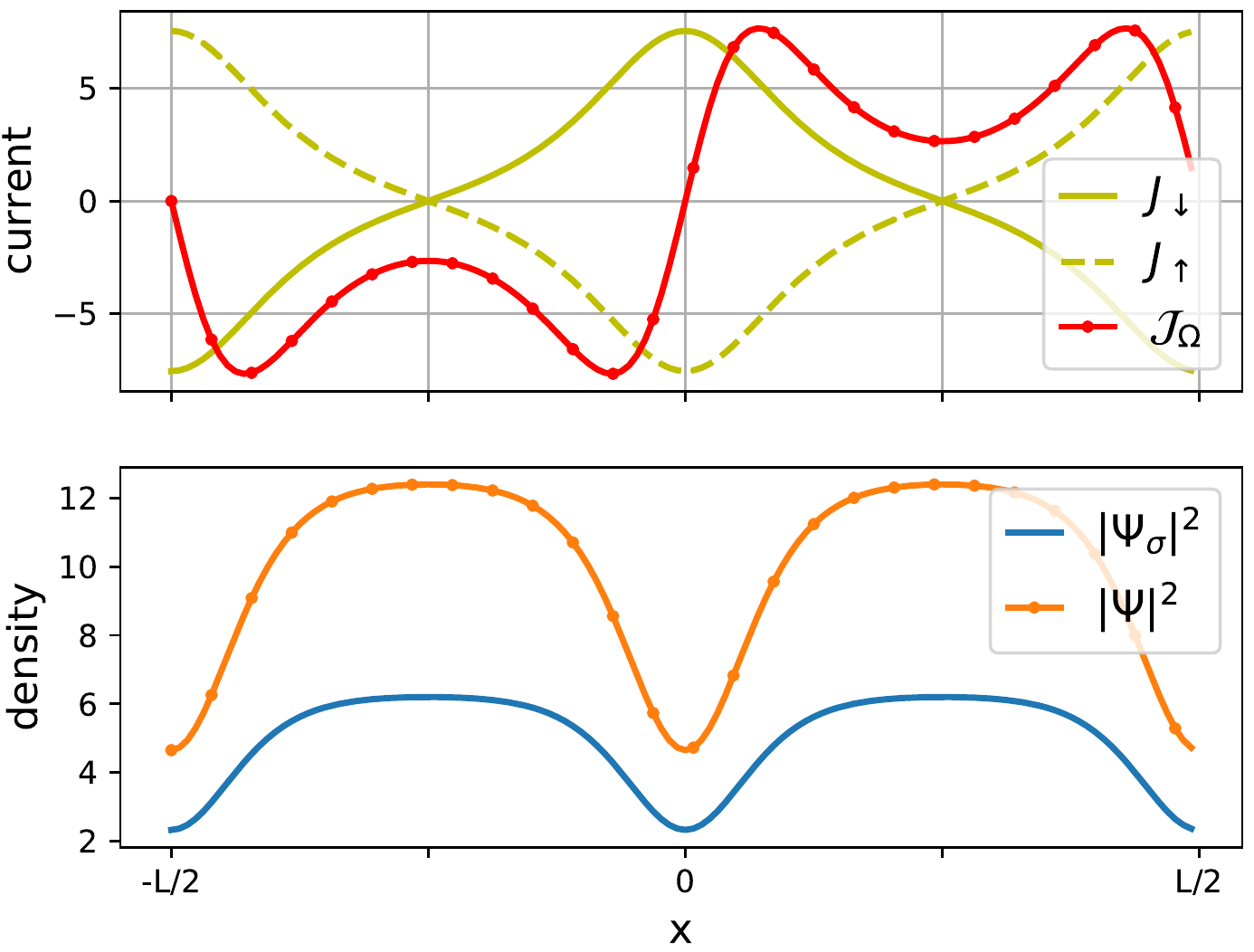} \hspace{1.4cm}
	\includegraphics[width=0.4\linewidth]{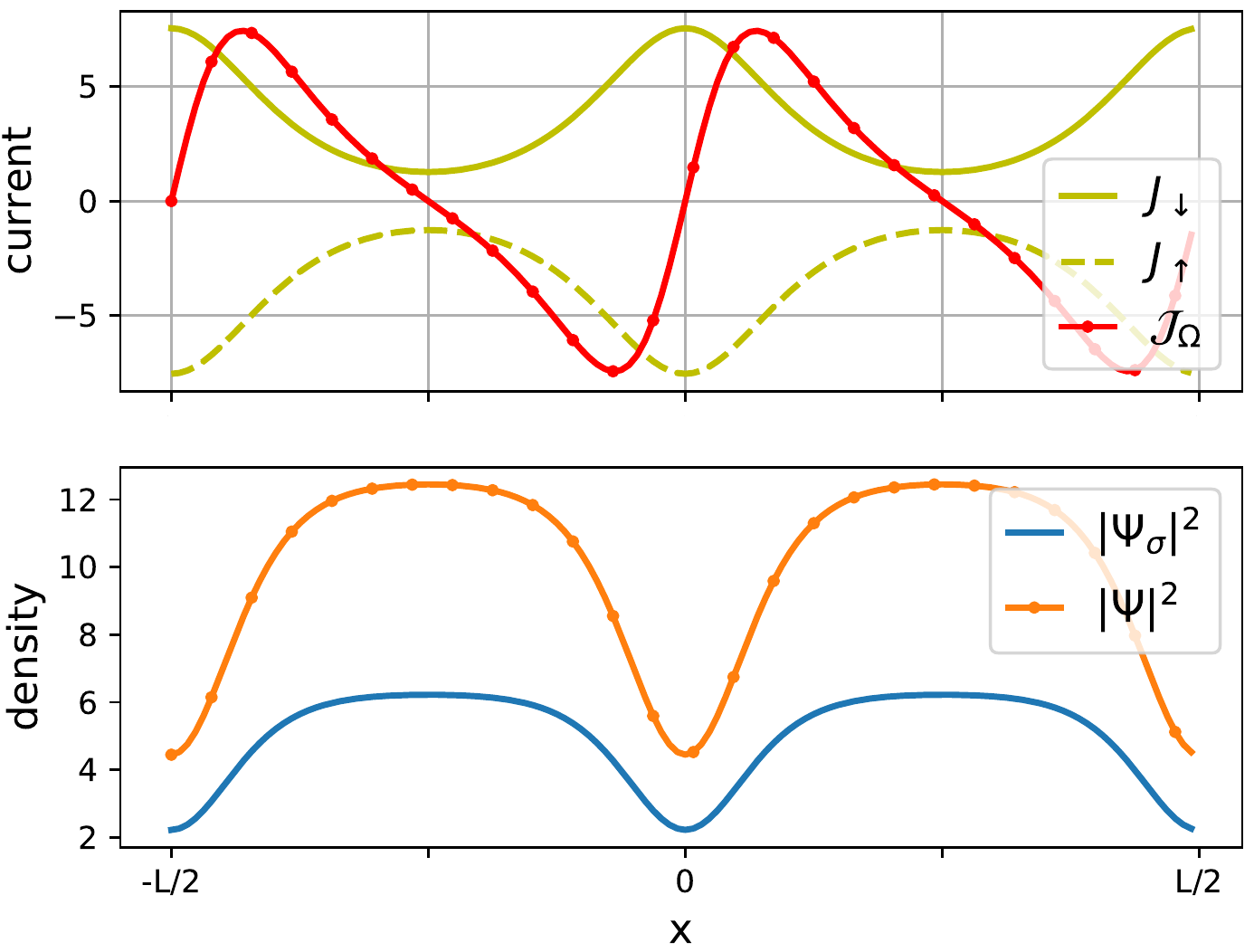}\\
	\vspace{-0.2cm}
	\flushleft (\textbf{c})  \hspace{1.05\columnwidth}(\textbf{d})\\
	\centering
	\begin{minipage}{\columnwidth}
		\centering
		\includegraphics[width=1.07\linewidth]{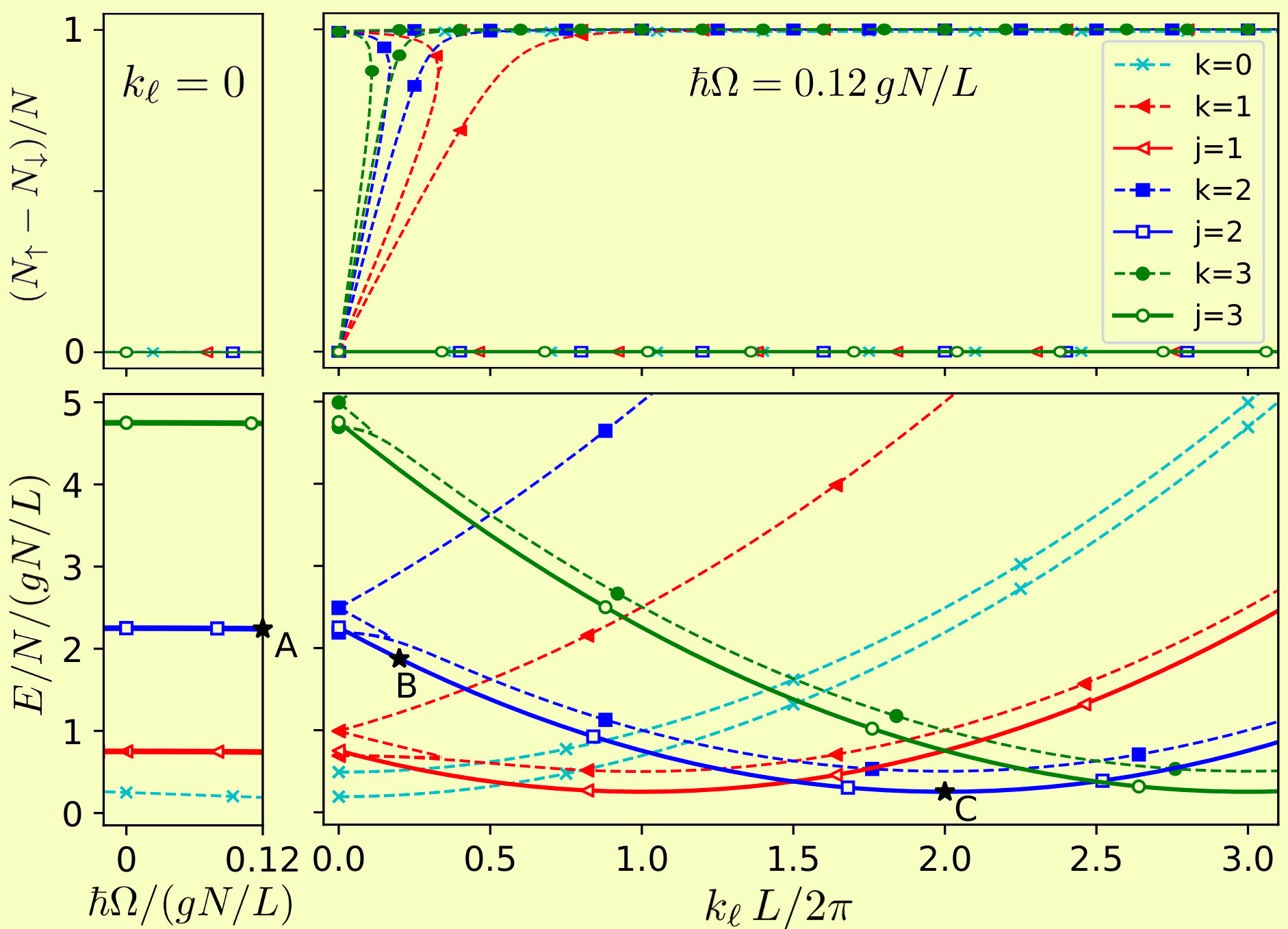} 
	\end{minipage}
	\hfill
	\begin{minipage}{\columnwidth}
		\centering
		\includegraphics[width=0.95\linewidth]{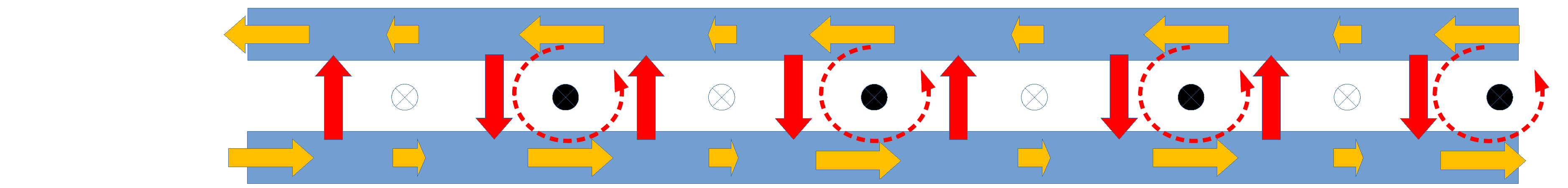}\\
		\includegraphics[width=0.95\linewidth]{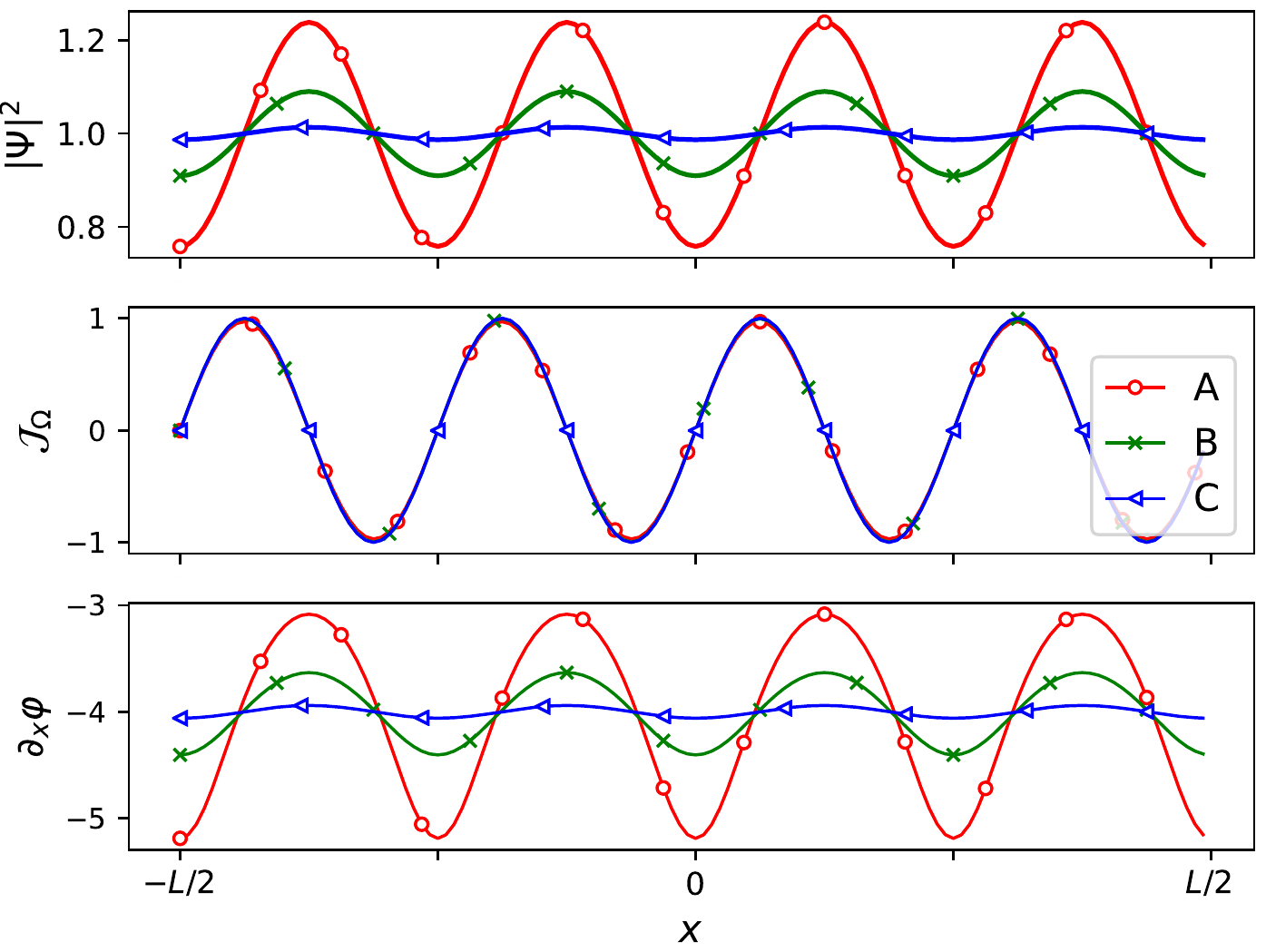}
	\end{minipage}
	\caption{(a)-(b) Counter-rotating (a) and coratating (b) Josephson-vortex arrays in a system with linear coupling $\hbar\Omega=0.2\,gN/L$ and particle number $gN/L=10\,\hbar^2(2\pi/L)^2/m$, in the absence of spin-orbit coupling. The schematic representation (top panels) indicates the vortex flows by dashed lines, the vortex cores by crossed-black circles, and other nodes in the currents by crossed-transparent circles.  The states show particle densities (bottom panels)  depleted at the vortex core positions, which in turn are surrounded by the vortex currents (middle panels).
		(c) Population imbalance (top panels) and average energy per particle (bottom panels) in plane wave states, labeled by the wavenumber $k$ in units of $k_0=2\pi/L$, and striped-density states, labeled by the number of pairs of coratating Josephson vortices $j$, in a system with $gN/L=\hbar^2(2\pi/L)^2/m$. In the absence of spin-orbit coupling $k_\ell=0$ (left panels), the striped-density states are excited-energy states for varying linear coupling $\Omega$. At fixed linear coupling $\Omega=0.12\,gN/L$  (right panels), such states are continued  in the spin-orbit-coupled regime ($k_\ell\neq 0$), where the striped density states become the system ground states in a coupling range centered at $2k_\ell=j\, k_0$.  The paths of the population-imbalanced plane waves are also shown for comparison (see the Appendix C for details). (d) Typical features of states with four coratating vortices, labeled A, B, C in panel (c), along a path of varying spin-orbit-coupling.}
	\label{fig:JV0}
\end{figure*} 
It turns out that neither the condensate current densities nor the Josephson currents depend locally on the density. 
As can be seen in the top panels of Fig. \ref{fig:JV0}(a), both group of solutions  describe junction-vortex arrays composed of $j=kL/4K(\mathfrak{m})$ vortices ($j/2$ vortex pairs due to the ring geometry), which are either counter-rotating, in Eqs. (\ref{eq:JV0_current}) and (\ref{eq:JV0_tunnel}), or coratating, in Eqs. (\ref{eq:JV1_current}) and (\ref{eq:JV1_tunnel}). In superconductors, these alternative configurations correspond, respectively, to a junction dynamics dominated by superconducting currents, or dominated by the applied magnetic field. In the neutral ultracold gas, the same picture essentially holds.

\section{ Analytical solutions}
The condensate densities $n_0(x)$ and $n_1(x)$ underlying the above junction-vortex arrays complement the  given particle currents to fulfill the GP Eq. (\ref{eq:gp}). We have found the corresponding analytical solutions to this equation that produce such densities in the absence of gauge, namely,
 \begin{equation}
 \Psi_{0}(x) = \sqrt{\bar n}\,\left[\,\binom{1}{1}\mathrm{sn}(kx,\mathfrak{m})+ \alpha \binom{\; i}{-i}\, \mathrm{dn}(kx,\mathfrak{m})\right],
 \label{eq:JV0}
 \end{equation}
 where $\alpha=\sqrt{1/\mathfrak{m}-2\hbar|\Omega|/g\bar n}$, and   $\bar n=\mathfrak{m} \{N/2L \,+ [1-\mathfrak{m}\,f(\mathfrak{m})]\,2\hbar|\Omega|/g \} $,  with $f(\mathfrak{m})=[K(\mathfrak{m})-E(\mathfrak{m})]/[\mathfrak{m}\,K(\mathfrak{m})]\, \in [1/2,\,1)$, and $E(\mathfrak{m})$ is the complete elliptic integral of second kind, 
and
\begin{equation}
\Psi_{1}(x) = \sqrt{\bar n}\,\left[\,\binom{1}{1}\mathrm{sn}(kx,\mathfrak{m})+ \alpha \binom{\; i}{-i}\, \mathrm{cn}(kx,\mathfrak{m})\right],
\label{eq:JV1}
\end{equation}
where $\alpha=\sqrt{1-2\hbar|\Omega|/g\bar n}$, and $\bar n=N/2L \,+ [1-f(\mathfrak{m})]\,2\hbar|\Omega|/g $. 
\begin{figure}[t]
	\flushleft (\textbf{a})  \\
	\centering
	\includegraphics[width=\linewidth]{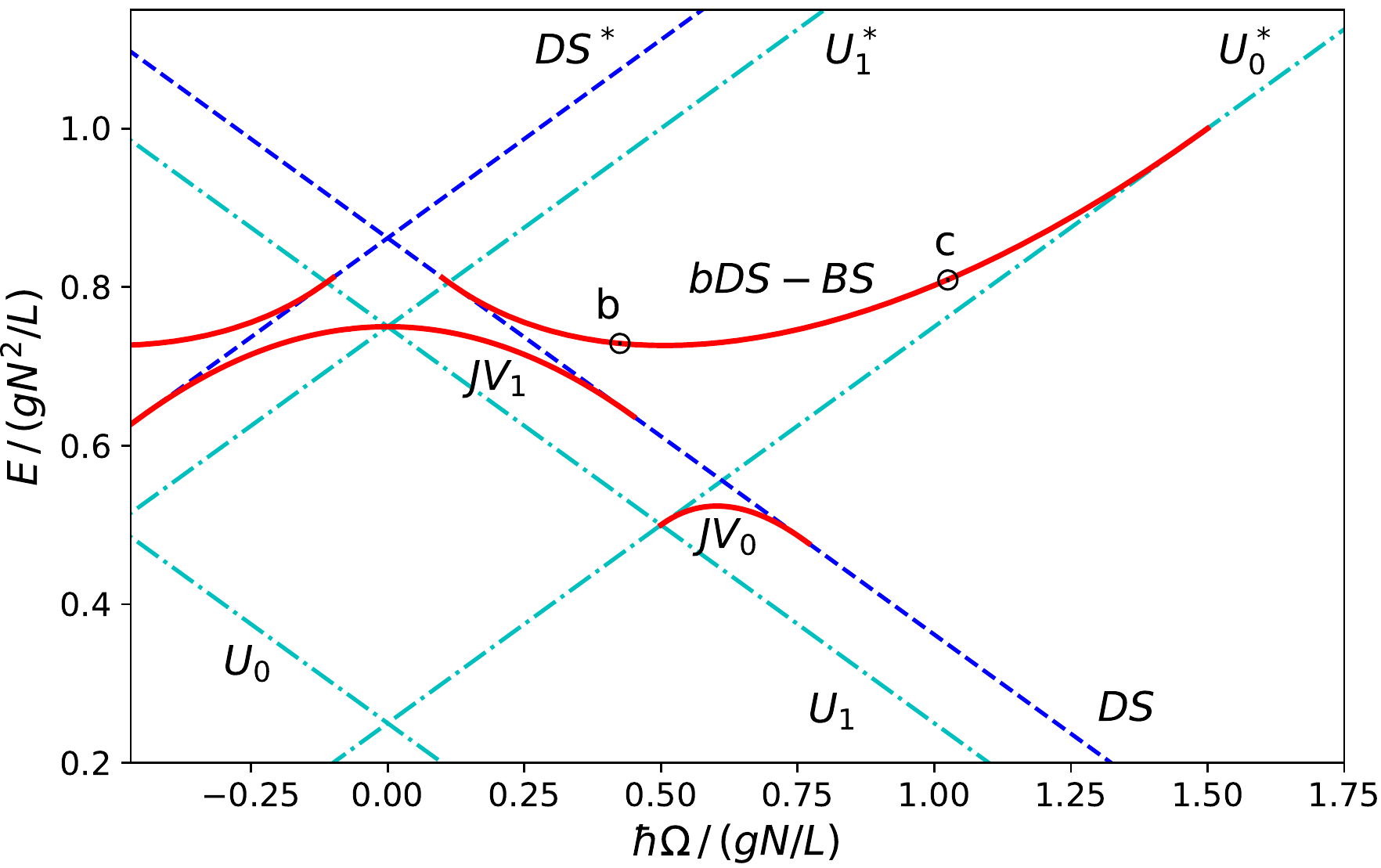}\\
	\vspace{-0.2cm}
	\hspace{0.2cm}
	\flushleft (\textbf{b})  \hspace{0.45\columnwidth}(\textbf{c}) \\
	\includegraphics[height=2.8cm]{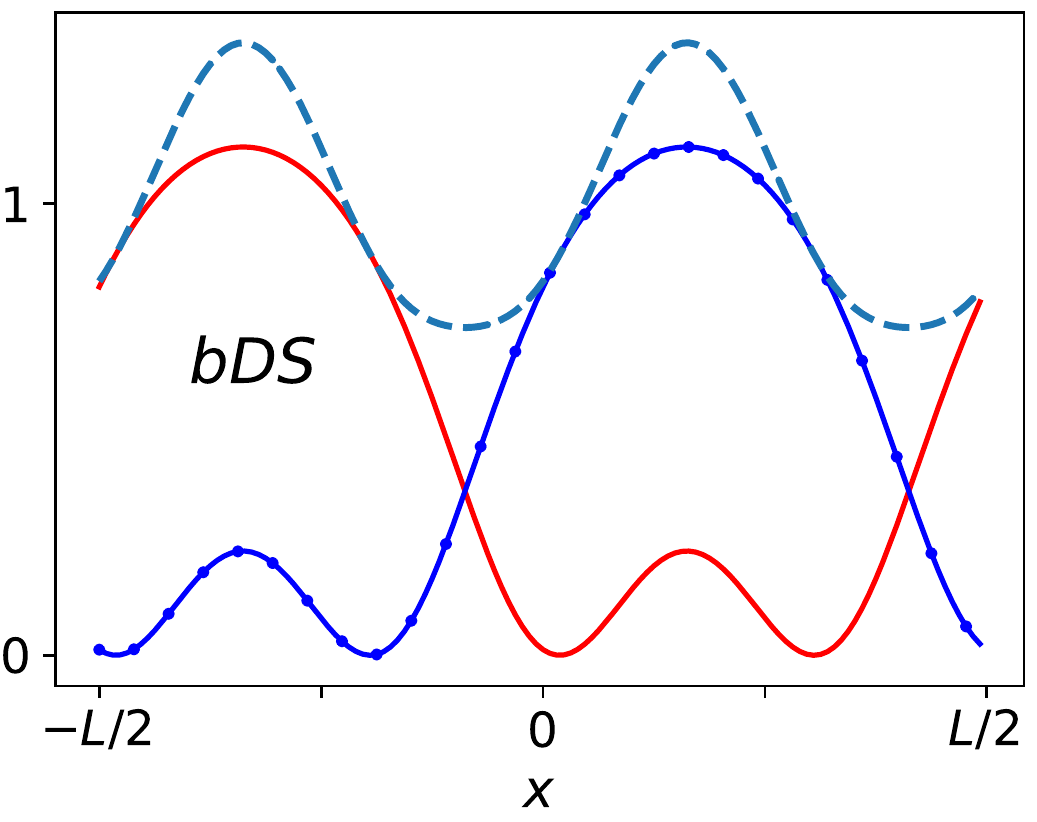}
	\hspace{0.05cm}
	\includegraphics[height=2.8cm]{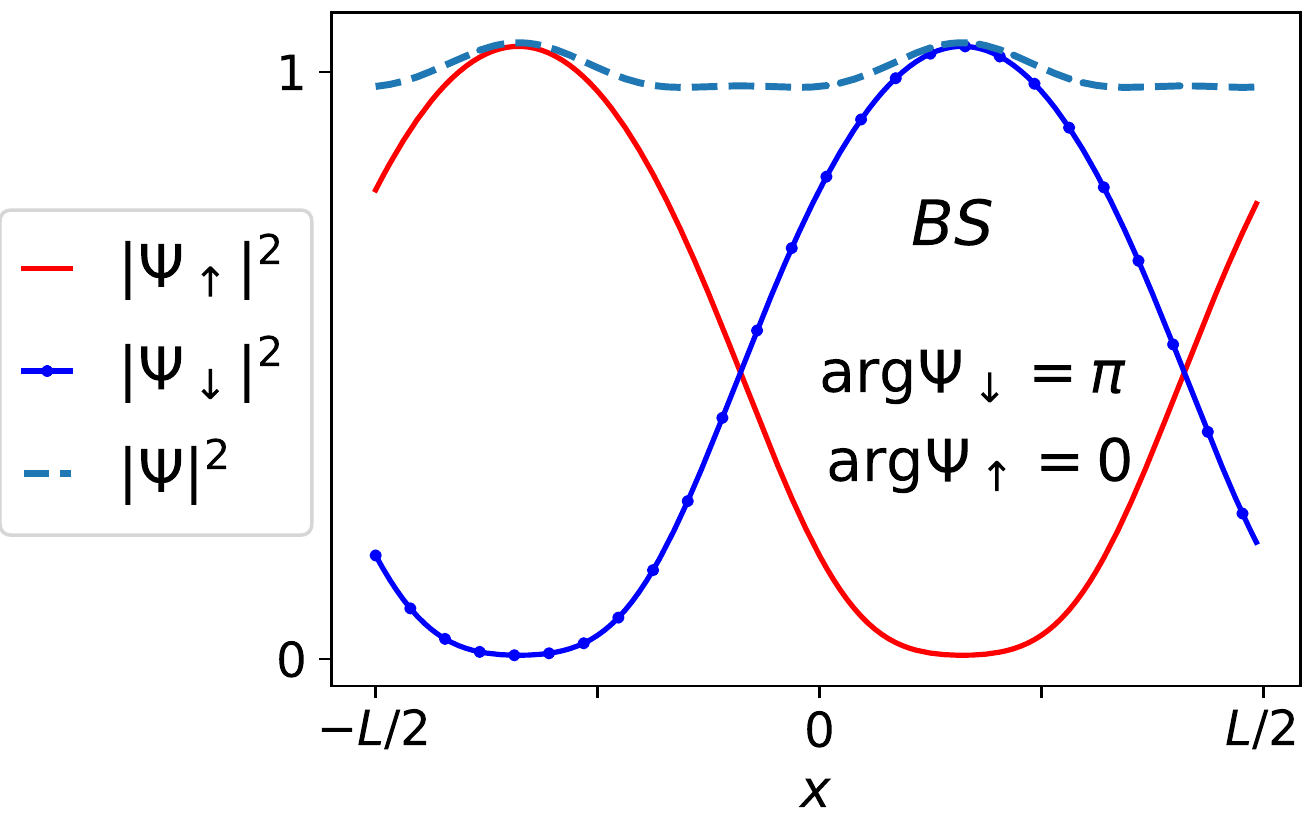}	
	\caption{(a) Energy per particle $E/N$ versus linear coupling $\Omega$ of parametrically connected stationary states with a fixed number of particles as given by $gN/L=\hbar^2(2\pi/L)^2/m$. 
	The continuous (red) lines represent coratating Josephson vortices JV$_{1}$, counter-rotating Josephson vortices JV$_{0}$, and bound-dark solitons along with bright solitons bDS-BS. The density profiles of states labeled $\mathbf{b}$ and $\mathbf{c}$ on the latter curve are shown in the bottom panels (b) and (c) respectively.
	Also depicted in panel (a) are the families of dark solitons DS and constant-density, uniform states U, in both in-phase and out of phase, denoted by a star superscript ($^*$), configurations. The states $U_0$ and $U_1$ have zero and minimum non-zero momentums respectively.
	 }
	\label{fig:connection}
\end{figure} 
By reversing the signs of the imaginary parts in Eqs. (\ref{eq:JV0}) and (\ref{eq:JV1}), energetically degenerate states with reversed chiral current are obtained.
The density contrast $\delta n=\{\max [n(x)]-\min [n(x)]\}/(N/2L)$ induced by the vortices is $\delta n_0=\mathfrak{m}\,2\hbar\Omega/(gN/2L)$ and $\delta n_1=2\hbar\Omega/(gN/2L)$, respectively. 
 A relevant difference between the two types of vortex arrays is the absence, for counter-rotating vortices (\ref{eq:JV0}), versus the presence, for coratating vortices (\ref{eq:JV1}), of total axial flows in each spin component. This feature allows the latter vortex array to find continuation into the regime of spin-orbit coupling, where net condensate currents are enforced by the gauge field. In this case, the degeneracy between states with opposite chiral currents is lost, and the state with current densities reduced by the gauge also acquires a reduced energy.
Due to the fact that in the limit of $\mathfrak{m}\rightarrow 1$, the elliptic functions show the asymptotic tendencies $\mathrm{sn}(kx,\mathfrak{m})\rightarrow \tanh(kx)$, $\mathrm{dn}(kx,\mathfrak{m})\rightarrow \mathrm{sech}(kx)$, and $\mathrm{cn}(kx,\mathfrak{m})\rightarrow \mathrm{sech}(kx)$, the
states (\ref{eq:JV0}) or (\ref{eq:JV1}) are the direct generalization to a vortex array of the single vortex solution given in Ref. \cite{Kaurov2005} (see also Ref. \cite{Aranson2002}).

The two junction-vortex families (\ref{eq:JV0}) and (\ref{eq:JV1}) have a simple interpretation as excited states
of the pseudo-spinor system, since they connect, by varying the linear coupling, distinct solutions of the GP Eq. (\ref{eq:gp}) that lack Josephson currents. To make the picture clearer, Fig. \ref{fig:connection} depicts these connections between families of stationary states in an energy-coupling chart for a system with fixed average density $gN/L=\hbar^2(2\pi/L)^2/m$ (see the Appendixes for the explicit energy expressions). 
 The families of in-phase (DS) and out of phase (DS$^*$) overlapped dark solitons (replicating the soliton trains in scalar condensates \cite{Carr2000}) are linked by families of coratating junction-vortex arrays  (JV$_{1}$) described by Eq. (\ref{eq:JV1}); the link is apparent when the energy-coupling graph is extended into the negative coupling domain. 
 On the other hand, the families of counter-rotating junction-vortex arrays  (JV$_0$) described by Eq. (\ref{eq:JV0}) connect, at lower energy, the
 uniform, constant-density states (U$_{0}^*$) with the family of in-phase dark solitons.
 It is worth pointing out that this latter connection is also made at higher energy by a family of non-current states (bDS-BS) composed by trains of staggered solitons, whose typical density profiles are shown in the bottom panels of Fig. \ref{fig:connection}. They can be described at low linear coupling as bound dark solitons (bDS) characterized by two close axial $\pi$-phase jumps, and at higher coupling as bright solitons (BS) featured by flat axial phase profiles, in spite of the repulsive interparticle interaction (see Ref. \cite{Zhang2020} about the role of this configuration in the decay dynamics of the out of phase uniform states U$_0^*$).
 A key distinction between the non-current states (bDS-BS) and the arrays of Josephson vortices is the dynamical stability of the latter (see the Appendix for details).

As can be seen in the compared numerical solutions of Fig. \ref{fig:JV0}(d), when the gauge is present the coratating Josephson vortices continue existing with the same  Josephson current (\ref{eq:JV1_tunnel}) obtained in the absence of gauge. 
For given coupling $\Omega$, a fixed number of junction vortices, playing the same topological invariant role in the pseudo-spin geometrical space as the winding number in the angular coordinate space of the ring, uniquely determines the tunneling flow for varying gauge.
As a result, the chiral currents are simply shifted from Eq. (\ref{eq:JV1_current}) in a constant that depends on the gauge momentum $\hbar k_\ell$. 
The relative-phase gradient and the particle density are also modulated in a similar way around the same average values $\sim 2\pi j/L$ and $\sim N/L$, respectively, where $j$ is the number of vortex pairs; just the modulation amplitudes vary, decreasing along with the difference $\Delta=2k_\ell-2\pi j/L$ [see the path traced by states A, B, C in Fig. \ref{fig:JV0}(c)] for increasing gauge momentum. 
Hence, the relative superfluid velocity and the system energy decrease with $\Delta$. 
This effect may be better understood with the gauge field entering the Josephson current as $J_\Omega\propto \sin (\varphi'-2 k_\ell x)$. It is then clear that if the relative phase is locked by the local coupling phase $\varphi'\sim 2 k_\ell x$ the particle tunneling is canceled $J_\Omega\sim 0$, and the corresponding coupling energy $E_\Omega\sim-\hbar\Omega\,|\Psi_{\uparrow}| |\Psi_{\downarrow}|$ minimizes the system energy. In this gauge, the relative phase reads $\varphi(x)'=2k_\ell\, x+\arcsin[2\,\bar n\,\alpha {\mathrm{sn}(kx,\mathfrak{m})\, \mathrm{cn}(kx,\mathfrak{m})}/{n(x)}]$. 

Since the vortex-array is energetically favored by the gauge field, it eventually reaches the system ground state for a particular range of the spin-orbit-coupling strength, around the value $k_\ell=2\pi j/L$ [point C in Figs. \ref{fig:JV0}(c) and (d)]. At this point, the chiral current density is shifted by $-2\hbar k_\ell\alpha\bar n/m$ with respect to the value given by Eq. (\ref{eq:JV1_current}), and shows oscillations around a zero average value. 
 Departures from the gauge value $k_\ell=2\pi j/L$ increase the energy of the vortex array with 2$j$ vortices, which becomes a metastable excited state beyond $k_\ell\sim2\pi (j\pm 1/2)/L$. Since the superfluid nature of the flow is the underlying cause of such metastability, in a similar way as in usual persistent currents \cite{Eckel2014}, hysteresis is expected to appear, and dissipative events caused either by quantum tunneling or thermal fluctuations will be responsible for phase-slip phenomena between adjacent vortex arrays with different numbers of vortices.

\section{ Experimental prospects}
The direct observation of junction-vortex arrays in bosonic junctions and the measurement of their associated particle flows represents a realistic goal.
 Not too long after the first observation of Josephson-vortex cores in superconducting junctions \cite{Roditchev2015}, experiments in two elongated condensates of ultracold bosonic gases loaded in double-well potentials have already observed sine-Gordon solitons in the relative phase of the coupled atomic clouds \cite{Schweigler2017}. 
 As can be inferred from the above discussion, the observed solitons are the signature of existing chiral currents and Josephson vortices in the system. 
 This experimental achievement, which reflects the ability of tuning the linear coupling and directly imaging the relative phase, presents a promising scenario for the  manipulation of junction vortices, as already happens in superconductors \cite{Dremov2019}, and thus for the control of the associated interference patterns \cite{Barone1982} that could be useful in the search for atomtronic devices.
 
Differently from the Josephson vortex solutions in just linearly coupled
 BECs, the presence of spin-orbit coupling (at low $m\Omega/\hbar k_\ell^2$) ensures the 
 vortex array emergence as the system ground state. This fact sets the
 experimental arrangement free from phase imprinting procedures and possible related
 temperature issues. 
 Even in a regime where the density stripes were not showing a high contrast, the
 usual absorption imaging of the atomic cloud during ballistic expansion would reveal the relative phase profile of the vortex array in the interference fringes. The characteristic $2\pi$ relative phase of each Josephson vortex is translated into inclined, seemingly broken fringes
 around the vortex position (see, e.g., Fig. 4 from the experiment in Ref. \cite{Schweigler2017},
 and Fig. 3 in Ref. \cite{Kaurov2006} from a theoretical description of the interference).
  We acknowledge likely difficulties for the experimental implementation of the general gauge field used in our model, having a non $2\pi$-periodic gauge in a ring geometry, but the particular periodic gauge, as well as a linear, non-periodic geometry, seem feasible targets. 
Although the presence of spin-orbit coupling produces the junction vortices at low linear coupling, it is not, as we have shown, a necessary ingredient. Alternatively, the excitation of relative currents between spin components, e.g., in a similar setting as  the toroidal spinor of Ref. \cite{Beattie2013}, could also trigger the emergence of the junction-vortex arrays (see also \cite{Roditchev2015} for a similar discussion in superconductors on the role of magnetic fields and relative superconducting currents).

\begin{acknowledgments}
This work was supported by the National Natural Science Foundation
of China (Grant No. 11402199), the Natural Science Foundation of Shaanxi Province(Grant No. 2018JM1050,
No. 2014JQ1022), and the Education Department Foundation of Shaanxi
Province(Grant No. 14JK1676).
\end{acknowledgments}

\appendix

\section*{Appendix A: Linear stability of junction-vortex arrays} 

\begin{figure}[t]
	\centering
	\includegraphics[width=\linewidth]{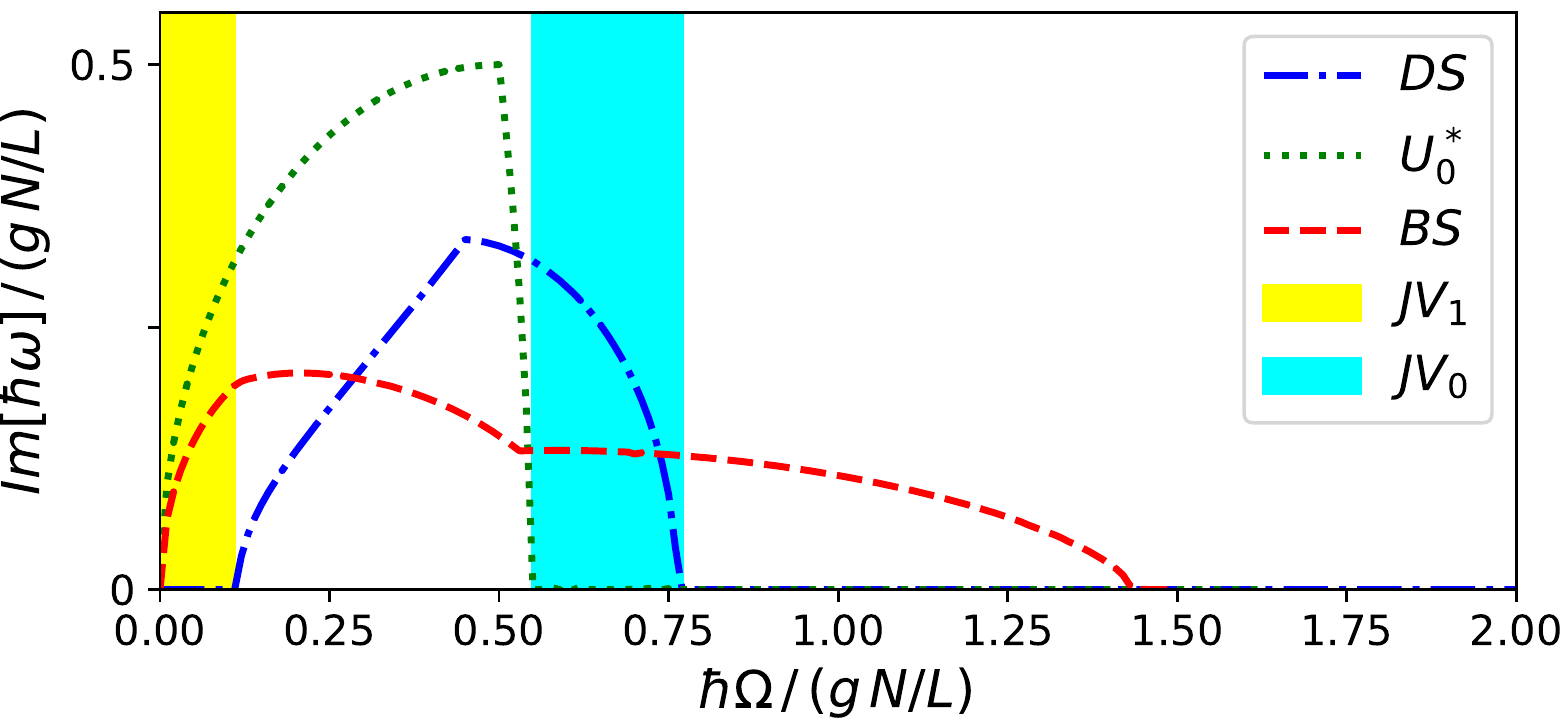}
	\caption{Energy of unstable linear excitations of the striped-density states represented in Fig. \ref{fig:connection}. The shaded regions correspond to ranges of dynamically stable junction-vortex arrays. These families also contain regions of instability connected to other unstable stationary states [overlapped dark solitons (DS) and out of phase uniform states (U$_0^*$) ]. Non-current states (labeled BS), made of out of phase staggered solitons, are dynamically unstable. }
	\label{fig:unstable}
\end{figure} 

The linear stability of stationary states $\Psi$ can be found from the Bogoliubov equations for the excitation modes $[u_{\sigma}(x), v_{\sigma}(x)]$ with 
energy $\mu+\hbar\,\omega$ \cite{Pitaevskii2003}:
\begin{equation}
\begin{aligned}
\left(-\frac{\hbar^2 \partial_x^2}{2m}+  2 g|\Psi_{\sigma}|^2 -\mu\right)  u_{\sigma}+   g  \Psi_{\sigma}^2  v_{\sigma} 
-\frac{\hbar\Omega}{2} u_{\bar\sigma}  = \hbar \omega \, u_{\sigma} \,,
\\
\left(\frac{\hbar^2 \partial_x^2}{2m}-  2 g |\Psi_{\sigma}|^2 +\mu\right)  v_{\sigma} -   g   (\Psi_{\sigma}^*)^2  u_{\sigma} 
+\frac{\hbar\Omega}{2}  v_{\bar\sigma} = \hbar \omega \, v_{\sigma} \,,
\end{aligned}
\label{eq:bog}
\end{equation}
where the subindexes \{$\sigma,\bar{\sigma}\}$ stand for $\{\uparrow,\downarrow\}$ and vice versa. Figure \ref{fig:unstable} reports our numerical results for the frequencies of unstable modes [hence with complex-frequency eigenvalues in Eq. (\ref{eq:bog})] for the families of stationary states with striped density profile (red lines)) presented in Fig. \ref{fig:connection}. As can be seen, only the families of junction-vortex arrays JV$_0$ and JV$_1$ contain dynamically stable configurations, showing the key role played by the Josephson currents as a stabilization feature in pseudo-spinor condensates. Even though unstable states also exist within these families, close to the connection points with other unstable stationary states, as overlapped dark solitons (DS) and out of phase uniform states (U$_0^*$).

\section*{Appendix B: Energy of junction-vortex arrays}

The energy functional for the pseudo-spin-$1/2$ system in a state $\Psi$ is
$E=\int dx[ \sum_{\sigma} ({\hbar^2}|(-i\partial_{x} +\sigma_z k_\ell)\Psi_{\sigma}|^2/{2m} + 
g \left\vert  \Psi_{\sigma} \right\vert ^{4}/2)-
{\hbar\Omega}\,\mathrm{Re}( \Psi_{\downarrow}^*\Psi_{\uparrow})]$.
Using this functional for the counter-rotating junction-vortex state (\ref{eq:JV0}) at $k_\ell=0$, the system mean energy is obtained as
\begin{align}
E_0=& {g\bar n^2L}\,  \biggl[ \alpha^4 + \frac{1-\mathfrak{m}\alpha^2}{3\mathfrak{m}}\left( 1+2\mathfrak{m}\alpha^2+\frac{3}{2}\alpha^2\right)  \nonumber \\  &
+f(\mathfrak{m})\frac{(1-\mathfrak{m}\alpha^2)^2}{3\mathfrak{m}}\left(\mathfrak{m}-\frac{1}{2}\right)\biggr] ,
\label{eq:JV0_energy}
\end{align}  
and the corresponding chemical potential is  $\mu= \hbar^2 k^2 (1+\mathfrak{m})/2m + g\bar n \alpha^2-\hbar\Omega/2$. 
This family exists for $\alpha\in[0,\,1/\sqrt{\mathfrak{m}}]$ and linear coupling values inside the range $\hbar\Omega\in[-g \bar n/2\mathfrak{m},\,g \bar n/2\mathfrak{m}\,]$. Notice that $ \bar n/\mathfrak{m}= N/2L \,+ [1-\mathfrak{m}\,f(\mathfrak{m})]\,2\hbar|\Omega|/g$ is also well defined for $\mathfrak{m}=0$.
For $\alpha=0$, at the maximum linear coupling $\hbar|\Omega|=g\bar n/2\mathfrak{m}$, the particle tunneling is suppressed, and the coratating vortex states merge with the family of dark-soliton trains, whose functional form is $\Psi_{DS}\propto \mathrm{sn}(kx,\mathfrak{m})$ \cite{Carr2000,Carr2000II,Kanamoto2009}. Within the interval $0<\alpha^2\le1/\mathfrak{m}$ the minimum linear coupling  is reached in the finite system when $\Omega=\hbar (2\pi/L)^2/2m$, and the coratating vortices transform into a uniform density state $\left\vert  \Psi_{\sigma} \right\vert ^{2}=N/2L$ without currents.

Analogously, for the coratating vortex arrays (\ref{eq:JV1}), the mean energy is given by
\begin{align}
E_1=&g\bar n^2 L\, \biggl[\frac{\alpha^2(1+\alpha^2)}{2}  \nonumber \\  &+ \frac{(1-\alpha^2)}{3\mathfrak{m}}\left( {1+2\alpha^2}+(1-\alpha^2)\frac{(2-\mathfrak{m})\,f(\mathfrak{m})}{2}\right)\biggr] ,
\label{eq:JV1_energy}
\end{align}  
and  the chemical potential is $\mu= \hbar^2 k^2 (1-\mathfrak{m})/2m + g\bar n-\hbar\Omega/2$. The family exists for $\alpha\in[0,1]$, and in the range of linear coupling $\Omega\in[-g\bar n/2,g\bar n/2]$.
For vanishing linear coupling $\Omega=0$, at $\alpha=1$, the system splits into two separated condensates with constant density $\bar n=N/2L$.
For $\alpha=0$, the coratating vortex states also merge with the family of dark-soliton trains.

\section*{Appendix C: Nonlinear plane-wave states}
The system of Eqs. (\ref{eq:gp}) admits plane wave solutions $\Psi=[\sqrt{n_\uparrow},\,\pm\sqrt{n_\downarrow}]^T\,\exp(ikx-i\mu_k t/\hbar)$, with total density $n=n_\uparrow+n_\downarrow$ and population imbalance $z=(n_\uparrow-n_\downarrow)/n\in [-1,1]$. It is convenient to parametrize the solutions in term of $\beta=\hbar\Omega/gn$ and $\zeta_k=2\hbar^2 k_\ell k/(mgn)$, so, for each wavenumber $k$, the chemical potential is given by $\mu_k=\hbar^2(k_\ell^2+k^2)/2m +gn/2\mp\hbar\Omega/(2\sqrt{1-z^2})$, and the imbalance is obtained from the solution of the quartic equation 
\begin{align}
z^4-2\zeta_k\, z^3-(1-\zeta_k^2-\beta^2)\,z^2+2\zeta_k\,z-\zeta_k^2=0.
\end{align}
Therefore, for given $\Omega$ and $k_\ell$, there exist four plane wave states for each wavenumber $k$. In particular, population balanced ($z=0$) plane waves ($k\neq 0$) are only possible for vanishing momentum shift $k_\ell=0$. 
In the limit $gn\rightarrow 0$, the imbalance tends to $z=\pm \zeta_k/\sqrt{\zeta_k^2+\beta^2}$, and so the chemical potential tends to the linear limit $\mu_k=\hbar^2(k_\ell^2+k^2)/2m\mp\sqrt{(\hbar^2 k_\ell k/m)^2+(\hbar\Omega/2)^2}$.

The average energy of the plane-wave states, as plotted in Fig. \ref{fig:connection}, is given by
\begin{align}
\frac{E_k}{N}=\frac{\hbar^2(k_\ell^2+k^2-2\,k_\ell\,k\,z)}{2m} + \frac{gn}{4}(1+z^2) \mp\frac{\hbar\Omega}{2} \sqrt{1-z^2}.
\end{align}
For high momentum shift $\zeta_k>1$, and low linear coupling $m\Omega/\hbar k_\ell^2<1$, the imbalance approaches $z\approx 1$, and the energy minimum, as a function of $k_l$, lies at $k_\ell\approx k$.

\bibliography{striped_density}

\end{document}